\documentstyle[11pt,aaspp4]{article}

\def\msun{$M_\odot$}

\def\etal{{\it et al. }}

\def\Mpc{{\rm Mpc}}

\def\beq{\begin{equation}}
\def\eeq{\end{equation}}
\def\beqar{\begin{eqnarray}}
\def\eeqar{\end{eqnarray}}
\def\msol{\hbox{$M_{\odot}$}}
\def\mtot{\hbox{$M_{\rm tot}$}}

\def\spesn{\hbox{$\varepsilon_{\rm SN}$}}

\def\vesc{\hbox{$v_{\rm esc}$}}
\def\avg#1{\langle {#1} \rangle}

\def\la{\mathrel{\mathpalette\fun <}}
\def\ga{\mathrel{\mathpalette\fun >}}
\def\fun#1#2{\lower3.6pt\vbox{\baselineskip0pt\lineskip.9pt
  \ialign{$\mathsurround=0pt#1\hfil##\hfil$\crcr#2\crcr\sim\crcr}}}

\def\iso#1#2{\hbox{${}^{#2}${\rm #1}}}
\def\he#1{\iso{He}{#1}}

\def\c1#1{\iso{C}{1#1}}
\def\n1#1{\iso{N}{1#1}}

\def\mej{m_{\rm ej}}
\def\avg#1{\langle #1 \rangle}
\def\sigbar{\avg{\sigma}}

\def\omegam{\Omega_{\rm WD}}
\def\omegab{\Omega_{\rm B}}

\def\rhob{\rho_{\rm B}}
\def\rhom{\rho_{\rm Macho}}
\def\fm{f_{\rm WD}}
\def\yld#1{{\cal Y}_{\rm #1}}

\def\lya{Ly$\alpha$}

\def\pcite#1{(\cite{#1})}
\def\pref#1{(\ref{#1})}

\def\bline{\rule[1.2mm]{3em}{0.1mm}}

\begin{document}

\slugcomment{submitted to {\it Astrophysical Journal}, 1999}

\title{
Chemical Abundance Constraints on White Dwarfs as
Halo Dark Matter}

\author{Brian D. Fields}
\affil{University of Illinois, Department of Astronomy \\
Urbana, IL 61801, USA \\
{\tt bdfields@uiuc.edu}}

\author{Katherine Freese}
\affil{University of Michigan, Department of Physics \\
Ann Arbor, Michigan 48109-1120, USA \\
{\tt freese@umich.edu}}

\author{David S. Graff}
\affil{Ohio State University, Department of Physics \\
Columbus, OH 42310-1168, USA
{\tt graff.25@osu.edu}}

\begin{abstract}
We examine the chemical abundance constraints
on a population of white dwarfs in the Halo of our Galaxy.  
We are motivated by  
microlensing experiments which have reported evidence
for massive compact halo objects (Machos) in the Halo
of our Galaxy,
with an estimated mass of $(0.1 - 1) \msol$;
the only conventional dark astrophysical candidates
for objects in this mass range are white dwarfs.  
However, our work constrains white dwarfs in the Halo regardless of
what the Machos are.
We focus on the composition of the material 
that would be ejected as the
white dwarfs are formed.
This material would bear the signatures of
nucleosyntheis processing, and contain
abundance patterns which can be used to
constrain white dwarf production scenarios.
Using both analytical and numerical chemical evolution models,
we confirm previous work that very strong constraints come
from Galactic Pop II and extragalactic carbon 
abundances.  We also point out that in some cases, depending
on the stellar model, significant nitrogen is produced rather
than carbon.  The combined constraints from carbon and nitrogen
give $\omegam h \la 2 \times 10^{-4}$ 
from comparison with the low abundances of these elements
measured in the Ly$\alpha$ forest.
We note, however, that these results
are subject to uncertainties regarding the nucleosynthetic yields of
low-metallicity stars. We thus investigate additional 
constraints from the light elements D and \he4,
the nucleosynthesis of which is less uncertain.
We find that these elements can be kept within observational
limits only for $\omegam \la 0.003$ and for a white dwarf
progenitor initial mass function sharply peaked at low mass
(2\msun).
Finally, we consider a Galactic wind, which is required to
remove the ejecta  accompanying white dwarf production from the galaxy.
We show that such a wind can be driven by Type Ia
supernovae arising from the white dwarfs themselves, 
but find that these supernovae also lead to unacceptably large abundances
of iron.  The only ways we know of to avoid these constraints
are that (1) the ejecta from low-metallicity Macho progenitors are
absent or competely unprocessed; or (2) 
the processed ejecta remain as hot ($\ga 0.3$ keV) gas which 
is segregated
from all observable neutral material to a precision of
$\ga 99 \%$.
Aside from these loopholes, we conclude 
that abundance constraints exclude white dwarfs as Machos.
\end{abstract}

\keywords{dark matter --- MACHOs}

\newpage

\section{Introduction}
\label{intro}

The nature of the dark matter in the haloes of galaxies is an
outstanding problem in astrophysics.  Over the last several
decades there has been great debate about whether this matter
is baryonic or must be exotic.  Many astronomers believed
that a stellar or substellar solution to this problem might be
the most simple and therefore most plausible explanation.
However, recent analysis of various data sets has shown
that faint stars and brown dwarfs probably constitute no more than
a few percent of the mass of our Galaxy (Bahcall, Flynn, Gould,
and Kirhakos \cite{bfgk}); Graff and Freese \cite{gf96a};
Graff and Freese \cite{gf96b}; Mera, Chabrier, and Schaeffer \cite{mcs};
Flynn, Gould, and Bahcall \cite{fgb};
Freese, Fields, and Graff \cite{freese}).  Hence the only surviving
stellar candidates of known populations are stellar remnants.
In this paper we consider severe constraints on white dwarf
stellar remnants.  The situation for neutron stars is probably
even more restrictive.  If indeed stellar candidates are ruled
out, one may be forced to more exotic nonbaryonic halo
dark matter.

We have been particularly motivated to consider white
dwarfs as Halo dark matter by recent results from
microlensing experiments (Alcock et al. \cite{macho:2yr};
Renault \cite{ren}), which have reported evidence
for Massive Compact Halo Objects (Machos) in the Halo of our
Galaxy. White dwarfs have been identified as plausible Macho
candidates because of the best-fit Macho mass of ($0.1-1$) \msun.
While some of our results are presented in the context of
a possible Macho interpretation, our chemical abundance
results constrain a white dwarf population in the Halo
regardless of what the Machos are.

In a previous paper (Fields, Freese, and Graff
\cite{ffg}), we discussed the baryonic mass budget
implied by a Galactic Halo interpretation of the LMC Macho events.
We found that a simple extrapolation of the Galactic
population (out to 50 kpc)
of Machos to cosmic scales gives a cosmic density 
$\rho_{\rm Macho} = (1-5) \times 10^9 f_{\rm gal} 
  \, h \, $\msun$ \, \Mpc^{-3}$,
which in terms of the critical density corresponds to 
\begin{equation}
\label{omegam}
\Omega_{\rm Macho}=(0.0036-0.017) h^{-1} f_{\rm gal} \, .
\end{equation}
Here the factor $f_{\rm gal} \geq 0.17$ 
is the fraction of galaxies that contain
Machos, as we argued in Fields, Freese, and Graff \cite{ffg},
and $h$ is the Hubble constant in units of 100 km s$^{-1}$ Mpc$^{-1}$.
This estimate applies regardless of the nature of the Machos,
and shows that Machos (if indeed they are in the Galactic Halo) are 
a significant fraction of all baryons.  Similar results
have been obtained by Steigman \& Tkachev \pcite{st}.

If one assumes--as we will hereafter--that the Machos are white dwarfs,
then stronger constraints result.
In particular, since white dwarfs are stellar remnants,
their formation necessarily requires both the formation of
progenitor stars, and ejection of the bulk of the progenitor mass
when the white dwarf is formed.
The simple requirement that the formation of white dwarfs is accompanied by
the release of at least as much mass in the form of hot gas
ejecta has profound consequences
which constrain white dwarfs as Machos.
For example, including progenitors in the Macho mass budget
increases the cosmological density of material needed to make Machos.
If Machos are white dwarfs resulting from a single
burst of star formation (without reprocessing of ejecta gas), then their
main sequence progenitors would have been 
at least twice more massive:
$\Omega_\star \geq (0.007 - 0.034) h^{-1} f_{\rm gal}$.  
Accounting for ejecta mass also has implications
on the scale of our Galaxy.  The gaseous ejecta produced
along with the Galaxy's Machos would have had a 
mass larger than what is measured in the known stellar
and gaseous components of the Galaxy.  Thus, mass budget
considerations demand that most of the ejecta
left the Galaxy, which in turn requires
some kind of Galactic wind to remove it.

The ejecta produced by the white dwarf progenitors
lead to constraints not only due to 
their mass, but also due to their composition.
The latter is the focus of this paper:
chemical abundance constraints on white dwarfs as Halo dark matter.
The ejecta contain the products of 
nucleosynthesis--enrichment of some elements,
depletion of others--which become signatures
of white dwarf formation.  We will show 
that current models for low-mass stellar nucleosynthesis 
predict a degree of processing which is so severe
that it rules out white dwarf Machos.

The most powerful constraints on white dwarfs as halo dark matter
come from carbon and nitrogen. However, the amount of these
produced is also dependent on the stellar model.  Hence
we also consider the less powerful but unavoidable constraints from the
light element abundances, deuterium and helium.
We find that \he4 can be kept within observational
limits only for the lowest possible Macho density 
$\Omega_{\rm Macho}$ compatible with Eq. 1, together
with high cosmic baryon density, and
Macho progenitor initial mass function (IMF) peaked at 2\msun (so that there
are very few progenitor stars heavier than 3\msun).  

The carbon and nitrogen yields 
from white dwarf progenitors depend on the IMF of the stars
and on the amount of Hot Bottom Burning, 
and are uncertain for zero metallicity stars.  Still, best
estimates for these yields are in excess of observations of these elements
in our Galaxy (as first discussed for the case of carbon
by Gibson and Mould (1997)).  Hence a galactic wind would
be required to eject these elements from the Galaxy
along with the excess mass.  We show that such a wind could be driven
by Type Ia supernovae, which are produced by the same
white dwarfs in binary orbits with other stars.  To produce
a successful wind, we find that at least
0.5\% (by mass) of stars must 
to explode as supernovae.  Such a scenario is reasonable,
since a comparable fraction of stars become supernovae in the Disk
of the Galaxy, if the star formation rate is
$\sim 1 \msol/{\rm yr}$ and the Type Ia rate is
$\sim 10^{-2}/{\rm yr}$ (Tutukov, Yungelson, \& Iben \cite{tyi}).
However, gas cooling may be
rapid enough to keep the bulk of the ejecta from being
evaporated.  Furthermore, even if the C and N are ejected
from the Galaxy, they are still constrained by extragalactic
observations.  Measurements of C and N in damped Lyman systems
and the Ly$\alpha$ forest are in excess of what would
be produced by a white dwarf Halo.
In addition, the Type Ia supernovae overproduce iron.

In Section \ref{IMF} we discuss white dwarf properties; we discuss
the initial mass function of the progenitor stars and the
relation between the masses of progenitor stars and the resultant
white dwarfs.  
In Section \ref{chemev}, we present our
chemical evolution models which calculate the 
effect of white dwarf production on
D, He, C, and O.
In Section \ref{sect:DHe}, we compare the expected
chemical abundances arising from white dwarf production
with observed D and He abundances
in various systems, and derive constraints on $\omegam$;
in section \ref{sect:CN} we derive constraints from
C and N, which in fact is more restrictive.
In Section \ref{wind},
we discuss the requirements for  a Galactic
wind to remove chemical debris from the Galaxy.
We finish with a discussion in Section \ref{conclude}.

\section{White Dwarf Properties: IMF and Initial/Final Mass Relation:}
\label{IMF}

{\it Initial Mass Function:}
The progenitor stars of any white dwarf halo had to arise from
an initial mass function (IMF) that is strikingly different from any
observationally inferred IMF: a white dwarf progenitor IMF must have
very few stars less massive than $\sim 1$ \msun, 
many intermediate mass stars, and few high mass 
stars with mass greater than $\sim 8$\msun.  
Adams and Laughlin (1996) 
argued that the initial masses of halo white dwarf progenitors have to
be between 1 and 8 M$_\odot$.  The lower limit on the range of
initial masses comes from the fact that stars with mass $< 1 M_\odot$
would still be on the main sequence.
The upper bound arises from
the fact that progenitor stars
heavier than $\sim 8 M_\odot$ explode as Type II supernovae, and leave behind
neutron stars rather than white dwarfs.
We can allow the IMF to have a small contribution
to higher masses so that there are some Type II supernovae and corresponding
remnant neutron stars, but not so many as to overproduce heavy elements.

Because low mass main sequence halo stars are intrinsically scarce
(Bahcall \etal \cite{bfgk}; Graff \& Freese 1996a,b),
an IMF of the usual Salpeter (1955)
type $dN/dm \propto m^{-2.35}$
is not appropriate, as it would imply a gross overabundance of low mass
stars in the Halo. Adams \& Laughlin (1996) propose a log-normal
mass function motivated by Adams \& Fatuzzo's (1996) theory of the IMF:
\begin{equation}
\label{lognormal}
\ln {dN \over dm}(\ln m) = A - {1 \over 2 \sigbar^2}
\Bigl\{ \ln \bigl[ m / m_C \bigr] \Bigr\}^2 \, . 
\end{equation}
The parameter $A$ sets the overall normalization. The mass scale
$m_C$ (which determines the center of the distribution) and the
effective width $\sigbar$ of the distribution are set by the
star-forming conditions which gave rise to the present day population of
remnants. Possible values of the parameters are $m_C=2.3 M_{\odot}$
and $\sigbar=0.44$, which imply warm, uniform star-forming conditions
for the progenitor population. These parameters saturate the twin constraints
required by the low-mass and high-mass tails of the IMF, as discussed
by Adams \& Laughlin (1996), i.e., this IMF is as wide as possible. 

Stars in the mass range 2-4 \msun
will produce different abundances of He, C, and N than an IMF with most
of the stars in the mass range 4-8 \msun.  
Thus we will also examine the effect of narrowly peaked 
IMFs chosen to highlight
the different nucleosynthesis within
the $1-8 \msol$ mass range. 

{\it Initial/Final Mass Relation:}
The relation between the mass of a progenitor star and 
the mass of its resultant white dwarf relies on an
(imperfect) understanding of mass loss from red giants.  We use
the results of 
Van den Hoek \& Groenewegen \pcite{vdhg}; these
are consistent with the results of Iben \& Tutukov \pcite{it}.
At the progenitor mass limits of interest, we have white dwarf masses
$m_{\rm WD}(1 $\msun$) = 0.55$ \msun, and
$m_{\rm WD}(8 $\msun$) = 1.2$ \msun.

\section{Chemical Evolution Calculations}
\label{chemev}

It is our goal to compare light element abundances
produced by white dwarf progenitors with the measurements
of the these abundances.  In this section we describe our approach to
evolution calculations to estimate the element abundances
arising from MACHO progenitors.
First, in Section \ref{sect:analytic}, we describe
two different extreme approximations to bracket the
possible abundances that can arise.  This analytic approach is
also useful in that it provides insight.  Then, in 
Section \ref{sect:numeric},
we discuss the numerical calculations.  Below, in Sections
4 and 5, we will apply these calculations to 
D and He, and then C and N.  There we will present
the results of our calculations and compare them
with observations of these elements.

Chemical evolution calculates the
history of gas as it is processed into stars,
which ultimately die, leaving remnants and ejecting
processed material.  Specifically, one calculates
the time development of the gas and comoving remnant
densities $\rho_{\rm gas}$ and $\rhom$,
as well as the gas density $\rho_{{\rm gas},i}$
in each isotope $i$.  The abundances $i$ are expressed
in terms of mass fractions $X_i = \rho_{{\rm gas},i}/\rho_{\rm gas}$.
All of these components 
change according to star formation and the resulting
star death.  
As initial conditions for all models, 
we take the baryons to be in gaseous form
with density $\rhob$.  We take the primordial composition
of elements to be the big bang nucleosynthesis abundances
appropriate for the chosen $\rhob$,
$X_i^0 = \rho_{{\rm gas},i}^0/\rho_{\rm gas}^0 = \rho_{{\rm gas},i}^0/\rhob$.
Here superscript $0$ refers to primordial abundances.

{\it Homogeneity:}
In both analytic and numerical calculations, we assume that at high
redshifts the gas exists in a single ``homogeneous" chemical phase;
i.e., concentrations of various element
abundances are independent of spatial position.
A corollary of this 
assumption is that outflow from stars is instantly and evenly mixed with 
the primordial gas.
This approximation allows us to use the average co-moving density of a 
chemical species as a useful parameter.  We will refer to $\rhob$ as the 
total co-moving baryon density, $\rho_g$ as the co-moving gas density, 
$\rho_{\rm WD}$ as the comoving white dwarf density,
$\rho_{\rm H}$ as the comoving hydrogen density, etc.
This picture thus amounts to a universal ``post-processing''
of baryons that occurs after primordial nucleosynthesis.

In reality some regions are likely to have abundances enhanced over the 
homogeneous levels, while other regions are likely to have abundances 
closer to primordial.  For example, the numerical simulations
of Cen and Ostriker (\cite{ceno}) suggest that the universe
is far from being chemically
homogeneous:  high density regions tend to have
higher metallicity than low density regions.
If mixing is less efficient than we have assumed, the element
abundances inside dense star forming galaxies due to progenitors
of white dwarf Machos
would be higher than our predictions, while the abundances
outside these regions would be lower.  
Lack of homogeneity makes the 
measured galactic abundances harder to match and
thus more constraining.  In the simulations
of Cen and Ostriker, the Ly$\alpha$ forest has a metallicity
roughly equal to the mean metallicity of the universe.
Thus, these forest lines are representative of the mean metallicity
results we calculate in our homogeneous models, and we
will use these lines below to compare theory with observation.
We do note, however, that a galactic wind
which drives material out of galaxies is likely to exist and
might be stronger than the one used in the Cen and Ostriker
simulations; such a wind drives the system towards homogeneity.
One can treat our results
as constraints on the efficiency with which the
enriched material is segregated from sites of subsequent
star formation.

\subsection{Abundances obtained with two Analytic Approximations}
\label{sect:analytic}

In this section we present analytic results of chemical abundances
obtained with two extreme approximations.
We consider two limits relating the star 
formation time-scale $t_{\rm SFR}$ to the lifetime of a typical 
star $t_*$ in  our strongly peaked IMF.  
In the limit where $t_{\rm SFR} \ll t_*$, or the {\it star burst} 
limit, all the Machos are formed in a short time.  Their ejecta mix into 
the IGM, but are not incorporated into any second generation of Machos.  
The opposite case where $t_{\rm SFR} \gg t_*$ is 
the {\it instantaneous recycling} limit.  
Here several generations of stars are 
created, and the ejecta from stars of one generation 
are mixed into the next generations of 
stars.  Within this limit, we can use the instantaneous recycling 
approximation of chemical evolution which ignores the lifetime of 
stars.  Note that a very efficient wind, which removes ejecta
into the IGM as soon as they are produced, would make the 
recycling case look more like a burst; in this case the ejecta from
a star are not mixed into the next generation of stars.
These two limits bracket any possible star formation scenario.

\subsubsection{Burst Model:}

We take the baryons in the universe at any time to consist of 
three components, with comoving densities:
\beq
\rhob = \rho_{\rm gas} + \rho_{\rm star} + \rho_{\rm WD} \, ,
\eeq
where subscripts ``star" and ``Macho" refer to stars and
remnant white dwarfs respectively.
Initially all the baryons are in gaseous form with different
primordial abundances of various species.  
During the star burst, a fraction $f_{\rm pro}$ of the
gas goes into stars, reducing $\rho_{\rm gas}$ 
from its initial density $\rhob$ by
an amount $f_{\rm pro} \rho_{\rm B}$.  Once the stars die, 
a fraction $R$
of the progenitor
mass is returned as processed gas.  
Given a white dwarf progenitor IMF
$\xi_*(m) = dN_*/dm$,  the gas return fraction is
\begin{equation}
\label{gasfraction}
R = {\int_{1M_{\odot}}^\infty dm \, \mej(m) \, \xi_*(m)
\over \int_0^\infty dm \, m \, \xi_*(m)} \, ,
\end{equation}
where $m$ is the mass of the progenitor,
which upon its death produces a remnant of mass $m_{\rm rem}$ 
and ejecta of mass $\mej = m - m_{\rm rem}$.
Thus, the density of ejected, processed gas is
$R f_{\rm pro} \rho_{\rm B}$; 
there is no further processing of the ejecta.
A portion of the progenitor stars is left in the form of white dwarf Machos.
These objects will have a cosmic density 
$\rho_{\rm WD} = f_{\rm pro} (1-R) \rhob$.
Thus a  ``white dwarf Macho fraction'' 
\beq
\fm \equiv \rho_{\rm WD}/\rhob = f_{\rm pro} (1-R)
\eeq
of the baryons is turned into white dwarfs.
Note that in the burst scenario, 
$\fm \le (1-R) < 1$.
In terms of the Macho fraction, the gas 
density 
after the burst is just 
$\rho_{\rm gas} = \rhob - \rhom = [1-f_{\rm pro} (1-R)] \rhob$
by baryon conservation,
and the gas 
fraction is $\mu = 1-\fm = 1 - f_{\rm pro}(1-R)$.
Hence, after the burst of star formation and the
evolution of the stars to stellar remnants has ended,
we are left with only gas and white dwarfs on the right hand
side of eqn. (3), with gas fraction $\mu$ and white dwarf
fraction $f_M$.

{\it Gas Composition:}
The initial gas density in each isotope $i$
is given by $\rho_{{\rm gas},i}^0 = X_i^0 \rhob$
where $X_i^0$ is the primordial abundance.  
As a result of star formation and the subsequent evolution
of the stars, the composition of the gas has changed to:
$\rho_{{\rm gas},i}  = \rho_{{\rm gas},i}^0 - f_{\rm pro} X_i^0 \rhob
+ \rho_i^{\rm eject}$.
The production of stars has lowered $\rho_{{\rm gas},i}$ by an amount
$f_{\rm pro} X_i^0 \rhob$. The ejecta of these stars once they die
has further changed it by $\rho_{{\rm gas},i}^{\rm eject}$.  The
details of this latter quantity depend on the element.
In the process of stellar evolution, some gas is turned into helium
and some primordial deuterium is destroyed.  In the remainder
of this section we describe our analysis of specific element
abundances in the burst model.

{\it Deuterium:}
All deuterium that passes through a star is destroyed.  
Thus, $\rho_{{\rm gas},D}^{\rm eject}
 = 0$, and 
the post-Macho D density is 
just that in unprocessed material:
$\rho_{{\rm gas},D} = (1 - f_{\rm pro}) X_D^0 \rhob$.
Thus the deuterium  mass
fraction $X_{\rm D}$ after the burst is
\begin{equation}
\label{Dburst}
X_{\rm D} = \frac{1-\fm/(1-R)}{1-\fm} \
   X_{\rm D}^0 \, .
\end{equation}

{\it Helium:}
As our notation we use $Y \equiv X_{\he4}$ to be the abundance
of \he4; we take the initial abundance to be $Y^0$. Some
of this helium is removed from the Galaxy by Machos, while
additional helium is added by the stellar evolution of the white dwarf progenitors.
In the case of helium, the ejecta are enriched:
$\rho_{{\rm gas},He}^{\rm eject}
 = ( Y^0 R + \yld{He}) f_{\rm pro} \rhob$,
where the first term is the fraction of the primordial helium that is returned
as processed gas after the stars die and the second term is
the He production during stellar evolution. The helium yield in the second term,
\begin{equation}
\label{eq:yld}
\yld{He} = 
  {\int_{1M_{\odot}}^\infty dm \, (m_{{\rm ej,He}} - Y^0 m_{\rm ej}) \, 
   \xi_*(m) \over \int_0^\infty dm \, m \, \xi_*(m)} \, ,
\end{equation}
measures the He production, over and above the initial abundance
$Y^0$.   Here $m_{\rm ej,He}$ is the mass of He ejected,
and $\mej$ is the total mass ejected.
For the Adams and Laughlin IMF (eq.\ \ref{lognormal}), and the Halo 
metallicity stellar yields of Van Den Hoek \& Groenewegen (1997), 
$\yld{He}=0.02$.  
Since the helium yield is a roughly constant function of mass,
$\yld{He}$ is roughly independent of IMF for a range of white dwarf 
IMFs.

The final, post-Macho He abundance is
thus $Y = (Y^0 \rhob - f_{\rm pro} Y^0 \rhob + 
   \rho_{{\rm gas,He}}^{\rm eject})/\rho_{\rm gas}$, which simplifies to 
\begin{equation}
\label{y}
\Delta Y = \frac{ \yld{He} }{1-R}  \ \frac{\fm}{1-\fm}
\end{equation}

{\it Carbon and Nitrogen:}
These elements have no primordial component, but are made by
stars.  Thus the production of C and N is formally
similar to that of He (eq.\ \ref{y}),
with the exception that the
lack of a primordial component means that
$X_{\rm C}^0 = X_{\rm N}^0 = 0$. 
Thus we have, after the burst,
\begin{eqnarray}
\label{cn}
X_{\rm C} & =  & \frac{ \yld{C} }{1-R}  \ \frac{\fm}{1-\fm} \\
X_{\rm N} & =  & \frac{ \yld{N} }{1-R}  \ \frac{\fm}{1-\fm} \, ,
\end{eqnarray}
where $\yld{C}$ and $\yld{N}$ are 
defined in a way analogous to eq.\ \pref{eq:yld}.

\subsubsection{Instantaneous Recycling Approximation}

Within the instantaneous recycling approximation (IRA), we have
the well known results (e.g., Tinsley \cite{tins})
\begin{eqnarray}
\label{recyc}
X_{\rm D} &=& ( 1-\fm )^{R/(1-R)} \ X_{\rm D}^0  \\
\Delta Y &=& \frac{\yld{He}}{1-R} \ln \frac{1}{1-\fm} \\
X_{\rm C} &=& \frac{\yld{C}}{1-R} \ln \frac{1}{1-\fm} \\
X_{\rm N} &=& \frac{\yld{N}}{1-R} \ln \frac{1}{1-\fm} \, .
\end{eqnarray}
Note that 
our ${\cal Y}_i \rightarrow (1-R) {\cal Y}_{{\rm Tins},i}$
in Tinsley's notation.
In this approximation there is no restriction on 
$\fm$, unlike the burst case (see below eqn. (5)).
Note also that as in the burst case,
the ratios $\Delta$He:C:N are constant.

The burst and recycling solutions 
agree to first order in $\fm$, but disagree at higher orders.
In particular, for a fixed $\fm$,
the burst model always gives a larger $\Delta Y$ and 
a smaller $X_D/X_D^0$ than the instantaneous
recycling approximation does.

\subsection{Numerical Models}
\label{sect:numeric}

The chemical evolution model used here
is based on  a code described in detail
elsewhere (Fields \& Olive \cite{fo98}).
The model allows for finite stellar ages
prior to the stellar death and the concomitant
remnant and ejecta production.
Thus the model assumes neither
instantaneous recycling nor the burst approximation,
which are equivalent to zero and infinite stellar lifetimes
respectively, relative to the timescale for star formation.
The star formation rate is chosen as an exponential
$\psi \propto e^{-t/\tau}$ with an $e$-folding time
$\tau = 0.1$ Gyr.  We have investigated other $e$-folding times
up to $\tau = 1$ Gyr
and find that the results are insensitive to details
of the star formation rate.
The initial mass function will vary as indicated.

The model results are only as reliable as the nucleosynthesis
yields used.  
For stars of $1-8 \msol$ we use the 
yields of Van den Hoek \& Groenewegen (1997),
which allow for metallicity-dependence (but the lowest
calculated metallicity is $Z=0.001$, i.e., 1/20 solar).
For higher mass stars we use the yields of Woosley \& Weaver \pcite{ww},
though the IMFs we examine put very little mass into these stars.

For the initial D and He abundances of our calculations, 
we have adopted the results of big bang 
nucleosynthesis calculations, which 
relate these quantities directly to $\rhob$ and the number of light 
neutrino species $N_\nu$.  We shall assume that $N_\nu=3$.  

As we will illustrate below, we find that our numerical calculations
yield results very similar to those of the burst approximation.
The reason for this similarity is that many of the stars are
in the low mass range, so that they have long lifetimes
compared to reasonable star formation rates.
By the time they die, they can no longer contribute to 
recycling in other stars.

\section{Deuterium and Helium}
\label{sect:DHe}

A large white dwarf component in the Galactic Halo
may lead to possible overproduction of helium and depletion
of deuterium.  The results of our calculations for these
two elements are presented in this section, and compared with
observations.  We will find that these elements can be kept
within observational limits only for $\Omega_{\rm WD} \leq 0.003$
and for a white dwarf progenitor initial mass function
sharply peaked at low mass (2\msun).

The problem of helium overproduction 
has previously been investigated by Ryu, Olive,
and Silk (1990).  In their work, they took the Galaxy to be 
a closed box, in which there is no infall of unprocessed 
gas to the Galaxy from the intergalactic medium (IGM), 
and no outflow of processed gas from the
Galaxy into the IGM.  They concluded that, in this closed box model,
the Halo could contain only a few white dwarfs, or else
the Galaxy would have no hydrogen left; all the hydrogen would have been turned
into helium.  We will generalize their work here: we will move
beyond the closed box model and
consider the possibility that the processed gas is able to leave
the Galaxy via a galactic wind.  The details of such a wind
will be discussed in a later section.

As we will see in Section 5, the overproduction of C and N
provide  by far the severest chemical abundance constraint on a white dwarf
population in the Halo.  However, this statement assumes that
we understand the dredge-up of C and N from the core
of the low-metallicity white dwarf progenitors (Chabrier
\cite{chabriernew}).  Hence, in this section
we consider D and He, whose yields are far less uncertain.
Of all of the elements considered here,
the evolution of D is the cleanest: D is always destroyed
by stars and is not produced in significant amounts
by any astrophysical process other than the big bang 
\pcite{els}.  
Although He is produced by stars, as are C and N,
He production is farther out from the core of the star
so that the He yields are thus less uncertain than those of C and N.
On the other hand, Fields \& Olive \pcite{fo98}
found that published He yields have trouble with the $Y-Z$ slope in
dwarf galaxies.  However, the difficulty was that
the model predictions {\em underestimate}
the slope compared to the observations, suggesting
that in fact the He yields themselves may be an underestimate.
In this sense, therefore, the constraints on He production
are conservative.

\subsection{Observational Constraints}

With the assumption of homogeneous abundances,
D and He are universally altered from their
primordial values.  In this view, then, 
the apparently ``primordial'' abundances
of D and He used to constrain BBN
are really ``pregalactic'' abundances
which have already had some
processing from their initial values.
We want to quote 
D and He abundances in different environments and use
these as constraints on processing by white dwarf progenitors.

{\it Deuterium:}
The best available Galactic measurement
of deuterium is the abundance in the present day 
local ISM. 
Linsky \pcite{linsky} find
${\rm D/H}=(1.5\pm 0.1) \times 10^{-5}$.
The present day value has been 
depleted by an unknown amount from 
the original low metallicity value by galactic 
disk stars, and thus provides a very conservative 
lower limit on the D abundance and thus on pre-Galactic processing.

A stronger limit arises from 
measurements of D in 
quasar absorption line systems.  
At present, different groups report different D/H values.
The strongest claims include
``high'' D/H $\simeq (8-25) \times 10^{-5}$ 
(Webb et al. \cite{webb}; 
Tytler et al.\ \cite{hityt})
measured in a system at
$z = 0.701$;
and ``low'' D/H $= (3-5) \times 10^{-5}$
(Burles \& Tytler \cite{bt98a}; Burles \& Tytler \cite{bt98b})
measured in two systems at $z > 3$.
These measurements are difficult and subject to systematic
errors (principally affecting H, rather than D).
It is thus unclear which (if either) of these values best
represents the primordial abundance.
Thus we will allowing a very generous range:
\beq
\label{eq:D}
{\rm D/H}_p = (3-25) \times 10^{-5} \, .
\eeq

{\it Helium:}
A best estimate of pre-galactic (i.e., normally
``primordial'') helium comes from extragalactic HII regions,
the lowest metallicity cases of which are in
blue compact dwarf galaxies.
The data are summarized in, e.g., Fields \& Olive \pcite{fo98}.
The large number of measurements now lead to a small statistical
error, so that {\em systematic} errors are now the limiting factor.
Again, we will take generous limits, adding the systematic error
linearly with the statistical errors (both at $1\sigma$):
\beq
\label{eq:He}
Y_p = 0.231 - 0.245
\eeq

\subsection{Model Results and Constraints}
\label{sect:DY-results} 

The results of our calculation 
depend on several parameters: the IMF of the white dwarf population,
the total density of white dwarfs
$\rho_{\rm WD}$, the Hubble constant,
and the total baryon density $\rhob$.  
In general, the departure 
from the big bang nucleosynthesis initial conditions increases as 
$f_{\rm WD} = \rho_{\rm WD} / \rhob$ increases, i.e., as 
white dwarfs become a larger fraction of the 
baryons.  We can see this in the analytical results.
As the white dwarf fraction increases
in Eqs. \ref{y} and eq. \ref{recyc},
helium and CNO enrichment increases, and
more deuterium is depleted. 

We present results for four different sets of parameter choices here.
In the first model, we take $\Omega_{\rm WD} h=0.0036$, the lowest value
allowed by a simple extrapolation of
the Galactic Macho results to a cosmic scale in Eq. (1)
(Fields, Freese, \& Graff \cite{ffg}).  In this  model
we take the white dwarf IMF of
Adams and Laughlin (eq. \ref{lognormal}).
Figure \ref{fig:std} summarizes the nucleosynthetic processing
in two panels.
In Figure \ref{fig:std}a, we show 
the values of $Y$ and D/H
which result from our calculations
(for various values of $\rhob$, and with $h=0.7$).
Shown are the
full numerical model, as well as the burst 
and instantaneous recycling models.  
Also shown are the initial values from big 
bang nucleosynthesis and the (very generous)
range of 
primordial values from eqs.\ \pref{eq:D} and \pref{eq:He}.
Note that the numerical model falls between the burst and
IRA, as expected.  It is interesting to see that the full model
falls very close to the burst case.  Thus we can conclude that
the burst model well-approximates the full results;
also, as the burst model gives stronger constraints,
the IRA results are in fact the most generous (and thus the most
conservative) bounds.

Since the previous model is obviously not consistent with measurements, we 
also present, in Figure \ref{fig:min_consis}, a threshold model with results
barely consistent with measurements 
of deuterium and helium.  For this 
model, we have kept the log-normal IMF suggested by Adam and Laughlin, but 
with different parameters: our IMF is centered at $M_c=2$\msun instead of 
$2.3$\msun, and is narrower, with an effective width $\sigma=0.05$ instead 
of 0.44. This IMF contains far fewer stars with initial mass $M>5$\msun, 
and so produces less helium enriched gas, represented by the fact that $R$ 
drops slightly from 0.69 to 0.66.  
We also drop $\Omega_{\rm WD} h$ down to 0.002, 
somewhat below the lower bound of what is
suggested by the simple extrapolation 
in eq. \ref{omegam} for $f_{gal}=1$.
This model is most constrained by the upper limit of the 
He data.  The allowed range in $\omegab$ is $0.01-0.03$ 
(for $h=0.7$).  Note that 
to prevent over-production of helium, Machos are 
a relatively modest $\sim 10\% $ of Baryons. 

Figures \ref{fig:min_imf2} and  \ref{fig:min_imf4}
represent the {\em minimum} cosmic processing
required if Machos are contained only in spiral Galaxies of
luminosities similar to the Milky Way:
$\omegam h = 6.1 \times 10^{-4}$ 
(Fields, Freese, \& Graff \cite{ffg}).
Figure \ref{fig:min_imf2} uses an IMF peaked at
$2 \msol$, designed
to minimize the effect on deuterium and helium abundances.
Figure \ref{fig:min_imf2}{\bf (a)} shows
that the effect on D and He is small and permissible
(but see the following section for discussion of 
C and N production in this model).
Figure \ref{fig:min_imf4} uses the same 
$\omegam$, but adopts an IMF peaked at 
$4 \msol$. Note the increased D and He processing now 
becomes unallowably large.  Thus we are driven to a low
initial progenitor mass by the helium and deuterium abundances alone.

Note that white dwarf progenitors would lead to a raised floor in the
\he4 abundance.  From Eq. (\ref{y}), one can see that, to obtain
the primordial helium abundance from the measured values,
one should really subtract the contribution due to white dwarf
progenitors.  This would complicate the usual big bang nucleosynthesis 
comparison of observed pregalactic abundances with
the primordial yields.

\section{Carbon and Nitrogen}
\label{sect:CN}

We illustrate here the difficulties of 
reconciling the carbon and nitrogen
production with the abundance of white dwarfs in the Halo
suggested by the microlensing experiments.

\subsection{Production of C and N}
\label{sect:CNprod}

White dwarf progenitors are expected to produce prodigious amounts
of C and N.  Here we discuss the relative production of these
two elements.
The relative amounts of C and N produced in
the asymptotic giant branch (AGB) phase are determined
by a process known as Hot Bottom Burning (hereafter HBB).  During
HBB, the temperature at the bottom of a star's convective
envelope is sufficiently high for nucleosynthesis to take place
(Sackmann et al \cite{sack}, Scalo \etal \cite{scalo2}, Lattanzio 
\cite{lattanzio}).
One of the main effects of HBB is to take the \c12 which is
dredged to the surface and process it into \n14 via the CN cycle.
Significant destruction of \c12 together with production
of \c13 and \n14 requires temperatures of at least
80 $\times 10^6$K. For low mass AGB stars ($m < 4$\msun), the effect of
HBB is negligible due to the low
temperature at the bottom of their envelopes.  For high mass AGB
stars (m $ >$ 4\msun), the effect of HBB depends on the amount of matter 
exposed to the high temperatures at the bottom of their envelopes,
the net result being the conversion of carbon and oxygen to nitrogen
(Boothroyd \etal \cite{booth}).
Yields of H, He, are not affected by HBB;
moreover, the total CNO yields also remain the same.
Since the CNO production is dominated by C and N,
this means that the sum C+N is independent of Hot Bottom Burning.
Thus, the main effect of Hot Bottom Burning is to
determine the degree to which C is processed into N,
but the sum remains the same.

With Hot Bottom Burning,
progenitor stars less massive than about 4 \msun\ produce significant
amounts of carbon and negligible nitrogen,
while heavier stars produce significant amounts of nitrogen and
negligible carbon.  
Van den Hoek \& Groenewegen (1997) find that a star of mass 
2.5\msun and metallicity $Z = 0.001$
will produce $1.76 $ \msun of ejecta of which $0.012 $ \msun is new
carbon, 
for an ejected mass fraction of $7 \times 10^{-3}$.  
In comparison, the solar system composition
has a carbon mass fraction of $3.0\times 10^{-3}$.  
In other words, the ejecta of a 
typical intermediate mass star have more than twice the 
solar enrichment of carbon.  
If a substantial fraction of all 
baryons pass through $1-4 \msol$ stars, the carbon abundance in this 
model will be near solar.  
These stars also produce $2.2 \times 10^{-4}$\msun of N,
leading to an ejected mass fraction
$1.25 \times 10^{-4} \simeq X_{\rm N,\odot}/8$,
a much lower enrichment.  On the other hand,
a 5\msun\ progenitor at the same metallicity
produces $X_{\rm C} = 7.2 \times 10^{-4} = 0.24X_{\rm C,\odot}$ and
$X_{\rm N} = 8.2 \times 10^{-3} = 7.4 X_{\rm N,\odot}$.
Hence, with Hot Bottom Burning,  a white dwarf
IMF with stars in the mass range 1-4 \msun produces a twice-solar 
enrichment of
carbon, whereas  a white dwarf IMF with stars in the mass range 4-8 \msun
produces seven times solar enrichment of nitrogen.  An 
IMF with stars in both regimes,
such as the Adams and Laughlin IMF in Eq. (2), produces both elements.

For comparison, van den Hoek and Groenewegen (1997) considered
the case of no HBB.  Then stellar yields of carbon are seen to
dominate the total CNO-yields over the entire mass range, with
C production at the level of solar enrichment.  
Models with HBB are favored as they are in excellent
agreement with observations, e.g. for AGB stars in the
Magellenic Clouds (Plez \etal \cite{plez}, Smith \etal \cite{smith}).
In the next section we will present results from our models
without Hot Bottom Burning; however, the presence of HBB
would not change our results as it merely trades a C overproduction
problem for a N overproduction problem.

A possible loophole to C and N overproduction 
stems from the primordial, zero-metallicity composition
that the Macho progenitors would have.
Stellar carbon and nitrogen yields for zero 
metallicity stars are quite uncertain,
and have not been systematically calculated
for the $1-8 \msol$ mass range of interest to us here.
Thus we use the yields of Van den Hoek \& Groenewegen (1997),
at the lowest metallicity, $Z=0.001 = Z_\odot/20$, and
as an approximation of the true $Z=0$ yields.
However, it is possible (although not likely) that carbon never
leaves the white dwarf progenitors, so that carbon overproduction is
not a problem (Chabrier \cite{chabriernew}).  Carbon is produced
exclusively in the stellar core.  In order to be ejected, carbon must
convect to the outer layers in the ``dredge up'' process.  Since
convection is less efficient in a zero metallicity star, it is
possible that no carbon would be ejected in a primordial star.  In
that case, it would be impossible to place limits on the density of
white dwarfs using carbon abundances.  
On the other hand, 
the 1\msun model of Fujimoto et al.\ \pcite{fkih}
suggests that C and N are in fact highly enriched
due to strong mixing.
Indeed, there is
evidence (Norris, Ryan, \& Beers \cite{nrb})
for very strong C enrichment in some Halo giants,
suggesting a mixing effect.

The basic result of typical models with HBB is then that
a white dwarf IMF with stars in the mass range 1-4 \msun produces a twice-solar 
enrichment of
carbon, whereas  a white dwarf IMF with stars in the mass range 4-8 \msun
produces seven times solar enrichment of nitrogen.  An 
IMF with stars in both regimes,
such as the Adams and Laughlin IMF in Eq. (2), produces both elements.
Without HBB, a solar enrichment of C is produced by all WD progenitor stars.

\subsection{Model Results}

In the figures, in panels b),  we show 
CNO abundances from the same four models discussed
previously for deuterium and helium.  The CNO abundances are presented
relative to solar via the usual notation of the form
\beq
[{\rm C/H}] = \log_{10} \frac{{\rm C/H}}{({\rm C/H})_\odot} \, .
\eeq
For example, in this notation $[{\rm C/H}]=0$ represents
a solar abundance of C, while $[{\rm C/H}]=-1$ is 1/10 solar,
etc.  Our C and N abundances were obtained without
including Hot Bottom Burning, which would exchange a C overproduction
problem for a N overproduction problem.  
The effect of HBB would be to increase N at the expense of C,
keeping the sum C+N constant.   

In Figure 1, we have $\Omega_{\rm WD} h = 0.0036$, the lowest
value allowed by Eq. (1). We take $h=0.7$ and the Adams-Laughlin
IMF in Eq. (2).  We see that, even after dilution with the primordial
baryons, 
the C and N abundances are still both greater than 1/10 solar
(e.g. [C/H] $>$ -0.8) over the entire range of $\Omega_B$.
Lower values of $\Omega_B$ correspond to higher C abundances
because there are fewer primordial baryons to dilute the C emerging
from the white dwarf progenitors.
In Figure 2, we have $\Omega_{\rm WD} h = 0.002$, $h=0.7$, and an IMF
peaked at 2\msun as described previously.  In Figures 3 and 4,
we have $\Omega_{\rm WD} h = 0.00061$, the minimum amount of WD
required to explain the microlensing results
if only Galaxies similar to ours produce WD Machos.
Figure 3 uses an IMF peaked at 2\msun while Figure 4 uses an
IMF peaked at 4\msun.  In all cases there is substantial
C and N production: in particular, the resultant C abundance is above
1/10 solar.

In the next section, we will show that, with or without HBB,
C and N exceed by at least 2 orders of magnitude the levels 
seen in halo stars in our own Galaxy as well as by an order
of magnitude those in quasar absorbers.  

\subsection{Observational Constraints} 

White dwarf progenitors produce a huge amount of C and/or N.
With the assumption of homogeneity, the C and N produced 
would give rise to a universal ``floor", i.e.,
an apparent Pop III component which might even be
mistaken as primordial.  If the abundances are not homogeneous,
then the observations of C and N in various sites 
can be used to obtain the required segregation of these
elements to keep them out of certain regions.  In addition,
if one argues that C and N are underrepresented in some region,
then they must be enhanced elsewhere.

The overproduction of carbon and nitrogen can be a serious problem,
as emphasized by Gibson \& Mould \pcite{gm}.
They noted that white dwarf progenitors are expected to be the
main source of carbon.  Thus the production of a white
dwarf population would be accompanied by a 
copious production of carbon, without a corresponding
enrichment of oxygen, which is made predominantly by
Type II supernovae.  The expected signature of
white dwarf production would be
anomalously high ratios of C/O and N/O,
i.e., ${\rm C/O} \ga 3 ({\rm C/O})_\odot$
and ${\rm N/O} \ga 3 ({\rm N/O})_\odot$.
However, metal-poor stars in our galactic halo have
C/O and N/O that are about 1/3 solar,
i.e., {\em below} and not above levels in Population I disk stars.
Thus Gibson \& Mould \pcite{gm} concluded that the gas 
which formed these stars cannot have been polluted by the ejecta of a 
large population of white dwarfs.  

In using Galactic Halo star abundance
ratios as constraints,
the Gibson \& Mould \pcite{gm}
analysis assumes that 1) the Halo stars form at the same time
as the white dwarf progenitors, and 2) the Galaxy's Macho progenitor
ejecta would remain {\em in situ}.   
It is possible that the observed low C spheroid 
stars formed {\it before} the white
dwarf progenitors, in which case they would not be affected
by the metals produced later on by the white dwarf progenitors.
The authors
note that galactic winds could intervene but argue
these to be unlikely.  However, they did not consider
the effect of Type Ia supernovae, which may in fact
be a natural engine to drive such winds 
(though at the price of iron production;
see \S\ref{wind}).  
Thus, in order to be generous to the
white dwarf scheme, we will examine C and N 
production in terms of the absolute abundances produced,
and use these as constraints on the degree of 
efficiency of the winds.

If the spheroid stars do not predate the white dwarf progenitors, then,
in our own Galaxy, the metal-poor Halo stars provide a strong constraint:
in these stars, neither C nor N has a 
detectable ``floor" that would indicate a pre-Galactic component.
However, there is no evidence for such a floor,
which would appear as a constant C and/or N abundance as,
e.g., Fe decreases.  
C has been observed with abundances at least as low as
$10^{-3} {\rm C/H}_\odot$;
and, N has been observed with abundances as low as 
$10^{-3} {\rm N/H}_\odot$.
Thus if the production of these elements is of order solar,
as we have seen in the previous section,
the segregation between white dwarf progenitor ejecta and these Halo
stars must be very effective.
Mixing must be prevented with a 
$\sim 99\%$ efficiency.
A way to achieve this segregation is with a Galactic wind,
which can remove C and N from the Galaxy.

If the C and N are expelled from the Galaxy,
the abundances of these
elements are constrained by measurements
in the intergalactic medium.
Carbon abundances in intermediate redshift 
\lya\ forest lines have been measured to be 
quite low.
Carbon is indeed present, but only at the
$\sim 10^{-2}$ solar level,
(Songaila \& Cowie \cite{sc}) in the  \lya\ forest at $z \sim 3$
with column densities $N \ge 3 \times 10^{15} \, {\rm cm}^{-2}$.
Ly$\alpha$ forest abundances have also been recently measured
at low redshifts with HST (Shull \etal \cite{shull}) to be
less than $3 \times 10^{-2}$ solar.

The forest lines sample the neutral intergalactic medium.
With HBB, white dwarf progenitors in the mass range ($1-4$)\msun
$\,$ typically produce solar abundances of carbon; without HBB,
all white dwarf progenitors do so.
If we assume that the nucleosynthesis products of
the white dwarf progenitors do not avoid the neutral medium,
then these observations 
offer strong constraints on scenarios for
ubiquitous white dwarf formation.
In order to maintain carbon abundances as low as $10^{-2}$ solar, only about 
$10^{-2}$ of all baryons can have passed through the intermediate mass 
stars that were the predecessors of Machos.  Such a fraction can barely
be accommodated by the results in our previous paper (Fields, Freese,
and Graff \cite{ffg})
for the remnant density predicted from our extrapolation 
of the Macho group results, and would be in conflict with
$\Omega_\star$ in the case of a single burst of star formation.
Note that, while the Halo star limit is not absolutely robust,
in that it could be avoided if the Halo stars predate the Machos,
the Ly$\alpha$ constraint cannot be avoided.  Hence, below,
in obtaining numbers, we use the Ly$\alpha$ constraint.

Furthermore, in 
an ensemble average of systems 
within the redshift interval $2.2 \le z \le 3.6$,
with lower column densities 
($10^{13.5} \, {\rm cm}^{-2} \le N \le 10^{14} \, {\rm cm}^{-2}$),
the mean C/H drops to $\sim 10^{-3.5}$ solar
(Lu, Sargent, Barlow, \& Rauch \cite{lsbr}).
One can immediately infer that, however carbon is produced
at high redshift, the sources do not enrich all material
uniformly.  Any carbon that {\em had} been produced more
uniformly prior to these observations (i.e., at still
higher redshift) cannot
have been made above the $10^{-3.5}$ solar level.
These damped Ly$\alpha$ systems are thought to be possible 
precursors to today's galaxies. 

While measurements of nitrogen abundance have not been
made in the Ly$\alpha$ forest, there are measurements
in damped Ly$\alpha$ systems.  The value of N/H in these systems
is measured to be typically $< 10^{-2}$ of solar,
and in one case at $z_{\rm DLA}=0.28443$ reported to
be as low as ${\rm N/H} = 10^{-3.79\pm0.08} {\rm N/H}_\odot$ 
(Lu et al \cite{lsbr}). In contrast, with HBB, white dwarf
progenitors in the mass range (4-8)\msun produce seven times
the solar abundance of nitrogen.
In order to reconcile measurements of C and N
in damped Lyman systems with the much higher abundances
predicted by white dwarf progenitors, one would have to argue
that these elements are ejected from the damped Ly$\alpha$ systems, which 
may be protogalaxies. Again a wind may be operative here.  However,
the segregation requirements are even stronger,
particularly if N/H of $10^{-4}$ solar is to be taken seriously.

{\it Comparison with Model Results:}
We can compare these observations with our model results to
obtain more quantitative constraints when specific parameter
choices are made.  Again, our models have no HBB included.
First let us assume that the abundances
we obtained in the figures apply homogeneously throughout the universe.
We will compare our results to the Ly$\alpha$ carbon measurements
of 10$^{-2}$ and the Halo measurements of $10^{-3}$.
Then in order to obtain agreement of 
the C and N abundances we find in our Model 1 (see Fig. 1)
with the Ly$\alpha$ observations described above (which are a factor of 30
below the predicted values),
we must reduce the white dwarf densities by a factor of 30.
Hence we require $\Omega_{\rm WD} h \leq 0.0036/30 = 1 \times 10^{-4}$.
Alternatively, we require an actual abundance distribution
that is quite heterogeneous: those regions in which the observations
are made must be underprocessed.  This implies departure from
the mean of a factor of at least 30, i.e., there must be segregation
efficiency of $1-1/30=97\%$.  

The other figures confirm the results of Figure 1.
While the parameter choices of Figures 2 and 3 give acceptably low
D and He reprocessing, the C and N abundances are again
10-100 times what is observed.  In Fig. 2 and 3, agreement with 
Ly$\alpha$ forest requires $\Omega_{\rm WD} h \leq 1 \times 10^{-4}$.
Figure 4, with
an IMF peaked at 4\msun, overproduces all four elements.
This last model is the least restrictive when comparing
with the Ly$\alpha$ measurements, $\Omega_{\rm WD} h \leq
2 \times 10^{-4}$.  Note that if C and N remain
inside the Galaxy and Halo stars do not
predate the white dwarf progenitors, then all these limits would
be an order of magnitude more powerful; the abundances must match
the measured C values of 10$^{-3}$ solar of the Halo stars.

Our results are mildly dependent on the redshift when C and N are expelled
into the IGM.  If the C and N are not expelled until low redshifts, then
they would not be seen in intermediate redshift $(z=2-3)$ absorbers.  Our
limits at low redshifts will be $\sim 3$ times
less restrictive since the observatonal limits are less restrictive.
However, removing the C and N from the Galaxy requires
supernovae.  Since
large numbers of SN Type Ia are not seen out to 
$z \sim 1$ (Hardin \etal \cite{hardin}), one must ensure that the 
supernovae have mostly gone off by $z \sim 1$.
Thus the stronger bounds quoted previously in the session
apply unless the supernovae that ejected the material
take place precisely at $z \sim (1-2)$.
Hence the low measurements of C and N in the damped Ly$\alpha$
systems are hard to reconcile with the higher
predictions of C and N from white dwarf progenitors.

Thus, C and N indeed prove to be very restrictive;
in {\em all} models the mean cosmic production is 
unacceptably large if it is homogeneously distributed.
As mentioned above, however, the abundances could well
be inhomogeneous due to galactic winds, 
which would blow the C, N, and other
products of the white dwarf progenitors out of galaxies.
The D, He, C, and N measurements could be avoided as constraints only if
there is not much mixing, e.g. of hot outflowing gas and cool
infalling gas; with mixing, the material
essentially reenters the galaxies with a universal proportion.

In summary, low mass stellar progenitors produce a solar enrichment
of carbon; high mass stellar progenitors produce either
a solar abundance of carbon (without HBB) or a ten times
solar enrichment of nitrogen (with HBB).  
Both elements are in conflict with measurements
inside our Galaxy and {\it must} be ejected from the Galaxy
if white dwarfs are to survive as Macho candidates.
Even outside our Galaxy, these abundances are hard to reconcile
with measurements of the Ly$\alpha$ systems.
We do wish to repeat the caveat, however, that the C and N
yields from low metallicity stars are still uncertain.

We close this section by
pointing out that extragalactic HII regions cannot contain
a substantial number of white dwarf Machos.  These regions are
observed to have N and C increasing as the oxygen abundance
increases.  White dwarf progenitors,
on the other hand, produce C and/or N without producing
O enrichment.  One would have to argue that extragalactic
HII regions missed out in white dwarf formation.

\section{Galactic Wind}
\label{wind}

We have seen that the progenitors of a substantial white dwarf Halo 
population would have produced a significant amount of pollution,
in conflict with observations.
In general one could avoid these constraints
by arguing for strong segregation between the hot gas emerging
from the progenitors and the cold gas where the element abundances
are measured.  Then one views the incompatibility of the
predicted abundances with the observations as a measure of the
required efficiency of segregation of the hot ejecta from
the rest of the universe.

A possible means of removing excess abundances from the Galaxy is a Galactic
wind.  As discussed in the Introduction, such
a wind is required to remove the excess gaseous baryonic material
left over from the Macho progenitors;  this excess material
has more mass than the Disk and Spheroid combined,
is extremely polluted (with carbon, nitrogen, etc.)
and must be ejected from the Galaxy. 
Indeed, as pointed out
by Fields, Mathews, \& Schramm \pcite{fms}, 
such a wind may be a virtue, as hot gas containing
metals is ubiquitous in the universe, seen in
galaxy clusters and groups, and present as an ionized
intergalactic medium that dominates the observed
neutral \lya\ forest.  Thus, it seems mandatory
that many galaxies do manage to shed hot, processed material.  
Here a galactic wind could remove helium, carbon and nitrogen 
from the star forming regions and mix it throughout the universe.  

Such a wind could be produced by supernova explosions providing
the energy source. The white dwarf IMF must therefore include
the stars responsible for the supernovae.
Possibilities include Type II supernovae from neutron stars
arising from massive progenitor stars; in this case the IMF
must contain some stars heavier than 8 \msun.  The disadvantage
of such a scenario is that these heavy stars evolve {\it{more}}
quickly than the lighter stars that give rise to the white dwarfs;
i.e., the supernovae explosions would naturally take place
before the white dwarf progenitors have produced their
polluting materials.  Then it would be hard to see how the
excess carbon and nitrogen could be ejected from the Galaxy.

We therefore propose the alternate possibility of Type Ia supernovae.
Here the same white dwarfs that are Macho candidates would
also be responsible for the supernova explosions.  These
white dwarfs are in binary systems.  Smecker \& Wyse (\cite{wyse})
have shown a problem with a binary system of two merging white dwarfs
as being responsible for the supernova explosions:  too few
such explosions are seen in haloes today to allow us to 
have enough of these earlier on to provide the required wind.
However, a scenario in which the white dwarf  has a red giant companion can
be quite successful.  The red giant 
loses mass onto the white dwarf. When the white dwarf
mass approaches the Chandrasekhar mass, then there is a supernova
explosion.  The timing is just right, since the supernova and 
accompanying galactic wind takes place when low mass stars become
red giants.  Thus the explosion and wind take place after the
white dwarf progenitors pollute the Galaxy with excess element
abundances, so that the wind is able to eject any excess helium, carbon and/or
nitrogen from the galaxy.  

Here we now show that about 0.5\% (by mass) of the stars
must explode as Type Ia supernovae in order
to provide sufficient energy to produce the required
Galactic wind.  Such a number is very reasonable, as it
is comparable to the number of Type Ia supernovae per white dwarf in the
disk of Galaxy.

Consider a protogalaxy with a baryonic mass
$M_B$, total mass $\mtot = M_B + M_{\rm DM} \sim 10^{12} \msol$,
and size $R \sim 100 \, {\rm kpc}$.
The escape velocity is thus
\beq
\label{eq:vesc}
\vesc^2 = 2 \ \frac{G \mtot}{R} \sim (300 \, {\rm km} \, {\rm s}^{-1})^2
\eeq
For a supernova wind to be effective in evaporating gas
from the protogalaxy, it must heat the gas
to a temperature $T_{\rm gas}$ such that the wind condition
\beq
\label{eq:evap}
\frac{3}{2} kT_{\rm gas} = \frac{1}{2} m_p v_{\rm gas}^2 
   > \frac{1}{2}m_p {\vesc^2}
\eeq
is satisfied,
or $kT_{\rm gas} \ga 0.3$ keV for the $\vesc$ value in
eq.\ \pref{eq:vesc}.

This condition sets a lower limit to the number (and fraction) of
supernovae needed, as follows.
We envision a scenario wherein some baryons (i.e., gas) become
stars and ultimately their remnants and refuse, 
while other gas remains
unprocessed.  We thus write
\beq
M_B = M_\star + M_{\rm unpro} \ \ ,
\eeq
and we will denote the ``processed fraction''
$f_\star = M_\star/M_B$.
Furthermore, we note that some of the white dwarfs
will occur in binaries and will lead to Type Ia supernovae.
Consequently, some (most) of the stars
will meet their demise as white dwarfs and planetary nebulae (PN), while
some will die as supernovae:
$M_\star = M_{\rm PN} + M_{\rm SN}$.
We thus
denote the ``supernova fraction'' 
$f_{\rm SN} = M_{\rm SN}/M_\star$;
our goal here is to constrain $f_{\rm SN}$.

To get the constraint, we assume that the
three gas components--unprocessed, planetary nebulae, and
supernova ejecta--are mixed, and come to some temperature
$T_{\rm gas}$.  Since the unprocessed and planetary nebula
components are much cooler than the supernova ejecta,
we can, to good approximation, put their temperatures to zero.
In this case, the temperature of the mixed gas is just
given by energy conservation:
\beq
\frac{3}{2} N_{\rm gas} \, kT_{\rm gas}  =  E_{\rm SN} N_{\rm SN}
\eeq
where $N_{\rm gas} = M_{\rm gas}/m_p$ is the number of gas molecules,
$N_{SN}$ is the number of supernovae that have gone off.
Also, $E_{\rm SN} \sim 10^{51} \, {\rm erg}$ is the mechanical
energy of the supernova, which is ultimately thermalized.
Furthermore, since $N_{\rm SN} = M_{\rm SN}/\avg{m_{\rm SN}}$, we have
\beq
\frac{3}{2} M_B \, kT_{\rm gas}  =  m_p \spesn M_{\rm SN}
\eeq
where $\spesn \equiv E_{\rm SN}/\avg{m_{\rm SN}}$ is the 
specific energy per supernova.  For Type Ia supernovae,
$\spesn \sim 10^{51} \, {\rm erg} / 5 \msol 
   = (3000 \, {\rm km} \, {\rm s}^{-1})^2$.

Collecting, then, we have
\beq
\frac{M_{\rm SN}}{M_B} 
   = \frac{3}{2} \frac{kT_{\rm gas}}{m_p \spesn}
\eeq
and since $M_{\rm SN}/M_B = f_{\rm SN} M_\star/M_B = f_{\rm SN} f_\star$, 
we have
\beq
f_{\rm SN} f_\star = \frac{3}{2} \frac{kT_{\rm gas}}{m_p \spesn}
\eeq
Thus the condition of eq.,\ (\ref{eq:evap}) gives
\beqar
f_{\rm SN} f_\star & > & \frac{1}{2} \frac{\vesc^2}{\spesn} \\
\Rightarrow f_{\rm SN} 
     & > & \frac{1}{2} \frac{\vesc^2}{\spesn} f_\star^{-1} \\ 
 & \sim & 5 \times 10^{-3} \ f_\star^{-1}
\eeqar
Thus we see that we need at least about 0.5\% (by mass) of the stars to 
explode as Type Ia supernovae; more, if the processed fraction $f_\star$
is significantly lower than unity.

Thus far, we have only accounted for gas heating due to
the Type Ia supernovae, ignoring any cooling processes.
However, cooling processes will operate; for the temperatures
of interest, the dominant cooling mechanism is bremsstrahlung.
We can estimate the importance of cooling by computing the
cooling rate, $\tau_{\rm cool} = E/\dot{E}$, where $E \sim kT \sim 0.3$ keV 
is the energy per gas particle, and $\dot{E}$ is the cooling rate
per particle.  The cooling rate is $\dot{E} = \Lambda n$,
with $\Lambda \simeq 10^{-23} \, {\rm erg}\, {\rm cm}^{3} \, {\rm s}^{-1}$,
and $n$ the gas density.  Assuming a constant density, we have
$n = {M_{\rm gas} \over {4\pi \over 3}  R^3}$, where $M_{\rm gas}$ and $R$
are the mass and radius respectively of the WD gaseous ejecta.  Thus
\beq
\tau_{\rm cool}  = 0.2 \ {\rm Gyr} \
   \left(  \frac{M_{\rm gas}}{10^{11} \msol} \right)^{-1} \ 
   \left(  \frac{R}{50 \, {\rm kpc}} \right)^{3}
\eeq
for the fiducial gas mass and radii indicated.
We see that the cooling timescale is shorter
than longest stellar lifetime considered, $\tau(2\msol) = 1$ Gyr.
Thus cooling can be effective if the Type Ia supernova burst is
not rapid or the WD progenitors have masses $\la 3 \msol$.   
Furthermore, the cooling will be all the more effective
if the gas is inhomogeneous, as denser regions will cool much faster.
On the other hand, the cooling is very sensitive to the assumed
total radius $R$ of the WD gaseous ejecta.  
Hence, cooling cannot rule out such a wind, but
it does demand that the wind be driven out on timescales
more rapid than $\sim 0.2$ Gyr.

Thus, if the cooling is indeed inefficient,
it is quite reasonable to use some of the white dwarf Macho
candidates as Type Ia supernovae to remove excess carbon
and nitrogen from the Galaxy.
However, SN Ia make prodigious amounts of iron, about
$m_{\rm ej}({\rm Fe}) \sim 1 \msol$ per event,
i.e., a large fraction of the mass going into Ia's
becomes iron
(Canal, R., Isern, J., \& Ruiz-Lapuente \cite{r-l}).
Thus we will expect a mass fraction of iron
of order 
\beq
X({\rm Fe}) \sim M_{\rm SN}/M_{\rm B} = f_{\star} f_{\rm SN}
    \sim 5 \times 10^{-3} \sim 4 \, X({\rm Fe})_\odot
\eeq
i.e., a very large enrichment.
Thus, while the SN Ia's can remove the gas
from the galaxies, they add their own contamination
which must be kept segregated from the 
observable neutral material at a high precision.
(And the iron makes things all the worse as it also
adds to the cooling of the hot gas.)

\section{Conclusions and Discussion}
\label{conclude}

In conclusion, we have found that the chemical abundance constraints
on white dwarfs as candidate Machos are formidable.
The D and \he4 production by the progenitors of white
dwarfs can be in agreement with observation for low
$\omegam$ and an IMF sharply peaked at low masses
$\sim 2$\msun.  Unless carbon is never dredged up from
the stellar core (as has been suggested by Chabrier \cite{chabriernew}), 
overproduction of carbon and/or nitrogen is problematic.
The relative amounts of these elements that is produced
depends on Hot Bottom Burning, but both elements are produced
at the level of at least solar enrichment.  Such enrichment
is in excess of what is observed in our Galaxy and must 
be removed.  A Galactic wind may have been driven by Type Ia supernovae,
which emerged from some of the same white dwarfs that are
the Machos.  However, Ly$\alpha$ measurements in the IGM
are extremely restrictive and imply that these elements
must somehow be kept out of damped Ly$\alpha$ systems.
In addition these Type Ia supernovae overproduce iron
(Canal, R., Isern, J., \& Ruiz-Lapuente \cite{r-l}).

In sum, there is no evidence in Galactic halo stars, in external
galaxies, or in quasar absorbers for the patterns of chemical
pollution that should be formed along with a massive population
of white dwarfs.  While this debris does carry the seeds of its
own removal in the form of Type Ia supernovae, the required
galactic winds must be effective in all protogalaxies, must arise
at redshifts $1 < z < 2$, and the debris must remain hot and segregated
from cooler neutral matter.  Given these requirements, 
we conclude that white dwarfs are very unlikely Macho candidates
unless they are formed in an unknown and unconventional manner.

With the failure of known stellar candidates as significant sources
of dark matter, one may be driven to exotic candidates. These
include Supersymmetric particles, axions, massive neutrinos,
primordial black holes (Carr \cite{carr}; Jedamzik \cite{jedam}) 
and mirror matter Machos (Mohapatra \cite{mohap}).

\bigskip
We thank Elisabeth Vangioni-Flam, 
Grant Mathews, Scott Burles, Joe Silk, Julien Devriendt, Michel Cass\'e,
Jim Truran, Nick Suntzeff, Sean Scully, and Dave Spergel for helpful
discussions.  We especially wish to thank
Dave Schramm, without whom none of us would be working
in the field of cosmology.
We are grateful for the
hospitality of the Aspen Center for Physics,
where part of this work was done.
DG acknowledges the financial support
of the French Ministry of Foreign Affairs' Bourse Chateaubriand
and the Physics and Astronomy Departments at Ohio State University.
KF acknowledges support from the DOE at the
University of Michigan.  
The work of BDF was
supported in part by
DoE grant DE-FG02-94ER-40823.

\newpage

\centerline{\bf FIGURE CAPTIONS}

\begin{enumerate}

\item
\label{fig:std}
{\bf (a)} The D/H abundances and helium mass fraction $Y$ for 
models with $\omegam h = 0.0036$, $h=0.7$, and 
the Adams-Laughlin IMF.  The red curves show the
changes in primordial D and He as a result of white dwarf production.
The solid red curve is
for the full chemical evolution model, the dotted red curve is
for instantaneous recycling, and the long-dashed red curve for the burst model.
The short-dashed blue curve shows the initial abundances; the error bars show
the range of D and He measurements.  We see that the processing drives
D and He out of the measured range. \\
{\bf (b)} CNO abundances produced in the same model as {\bf a}, here plotted
as a function of $\omegab$.  The C and N production in particular are
greater than 1/10 solar (e.g., [C/H]$>-0.8$) 
over the entire range of $\omegab$.  Thse models do not include Hot Bottom
Burning; the effect of Hot Bottom Burning would be to
increase N at the expense of C, keeping the sum C+N constant.

\item
\label{fig:min_consis}
As in Figure 1, for 
$\omegam h = 0.002$, $h=0.7$, and IMF peaked at $2 \msol$.  
This is the absolute largest $\Omega_{\rm WD}$
compatible with data for the light elements.

\item
\label{fig:min_imf2}
As in Figure 1, for 
$\omegam h = 0.00061$, $h=0.7$. 
This represents the {\em minimum} cosmic processing
required if Machos are contained only in spiral Galaxies of
luminosities similar to the Milky Way.
The IMF is peaked at
$2 \msol$, designed
to minimize the effect on abundances.
We see in {\bf (a)} that the effect on D and He is small and permissible,
but in {\bf (b)} we see that even here the C and N production is siginificant.

\item
\label{fig:min_imf4}
As in Figure 2, for 
$\omegam h = 0.00061$, $h=0.7$. To show the effect of the IMF choice,
here the IMF is peaked at 
$4 \msol$. Note the increased D and He processing now 
becomes unallowably large.

\end{enumerate}

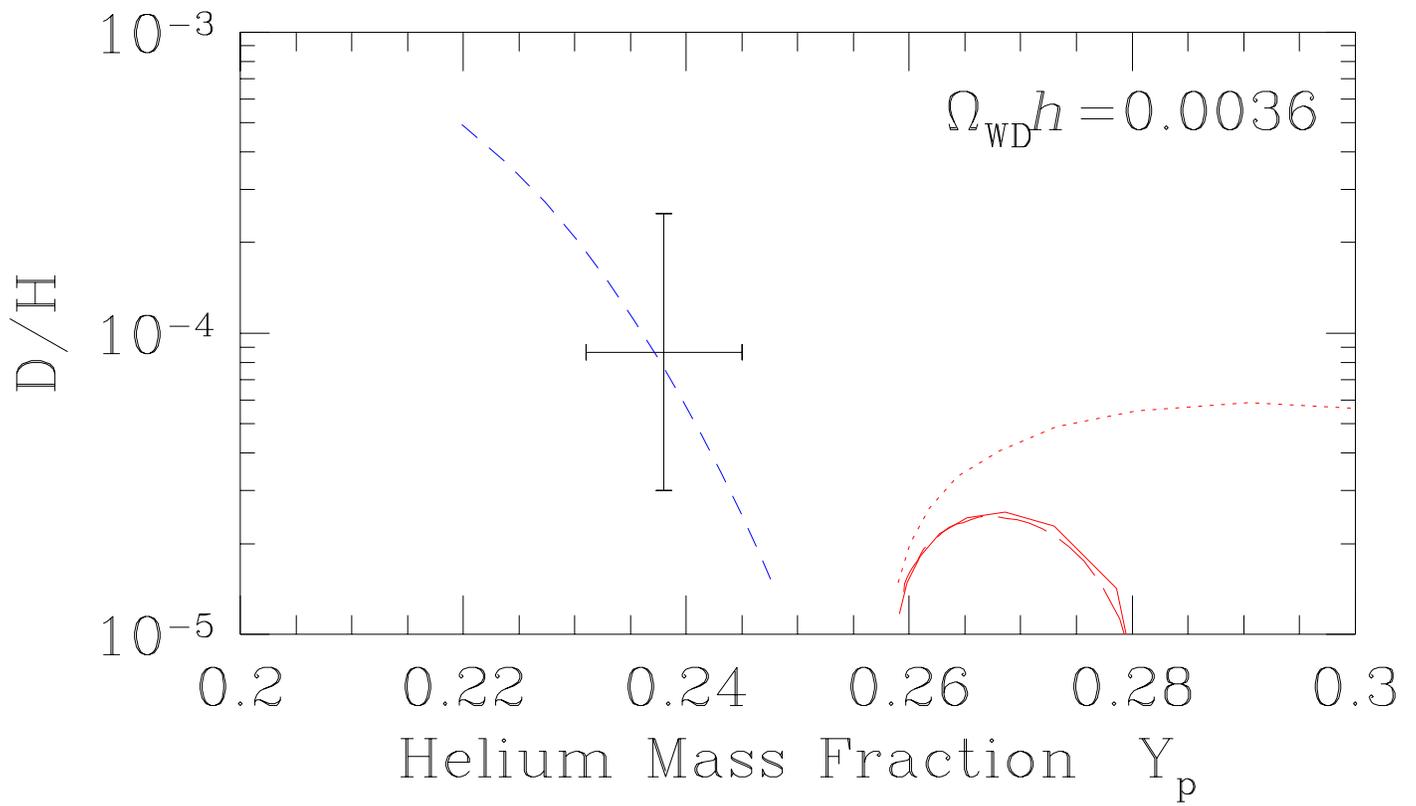
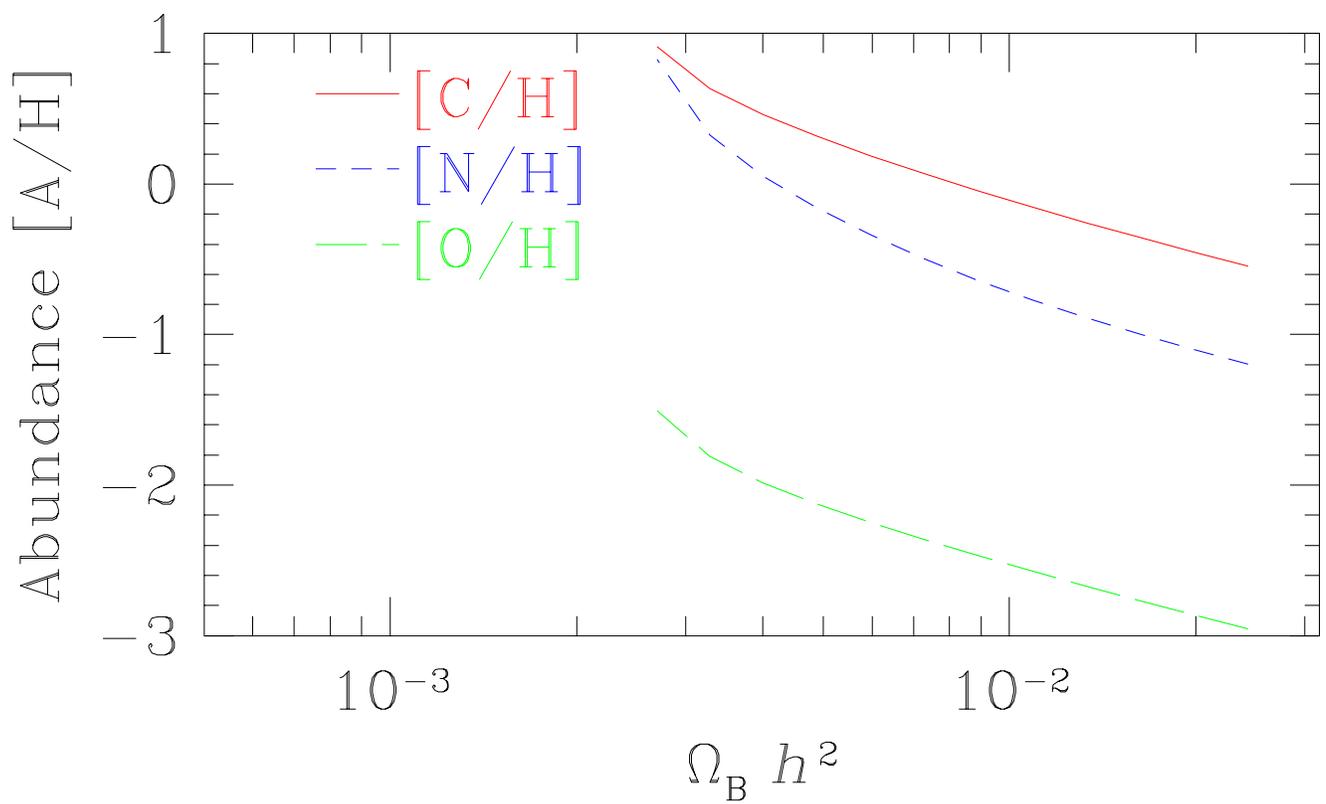
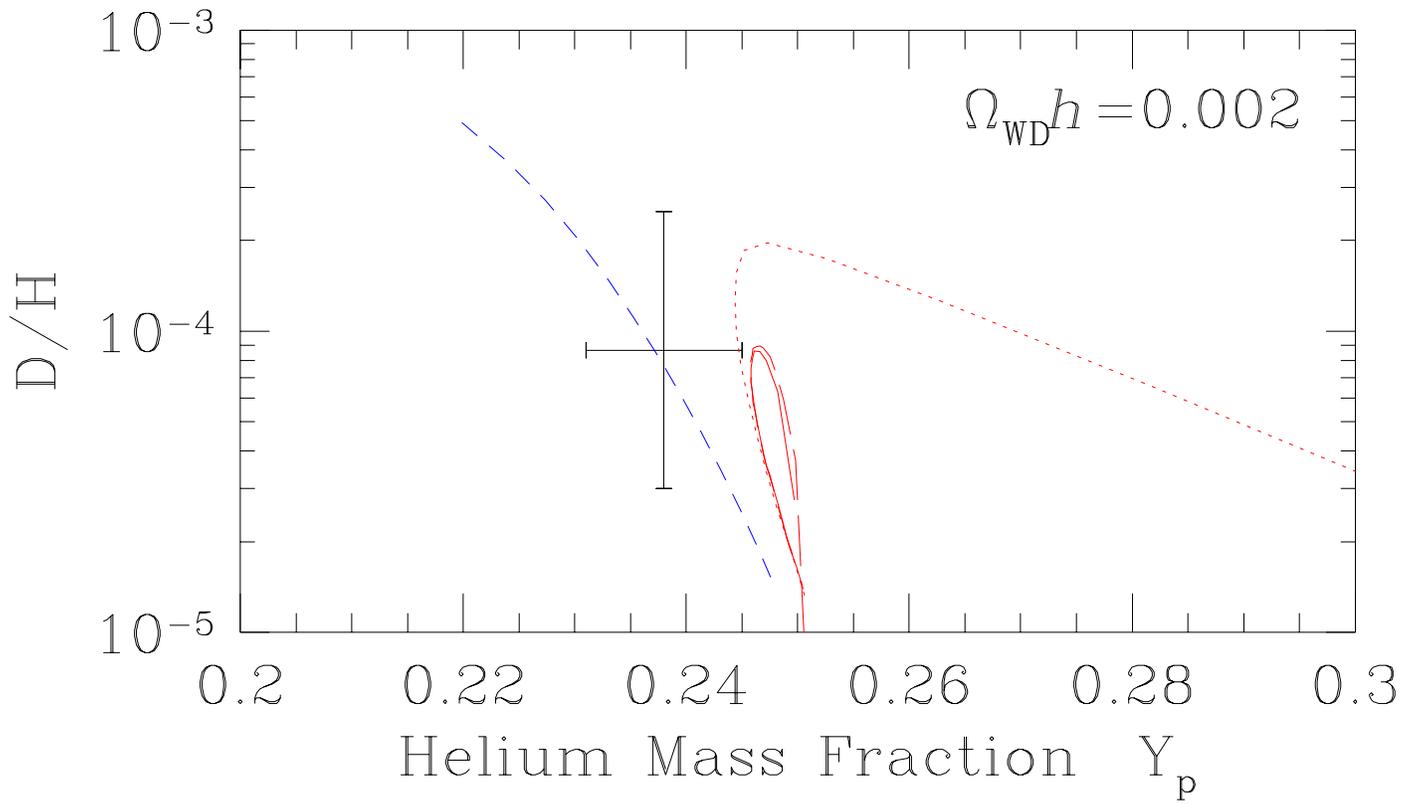
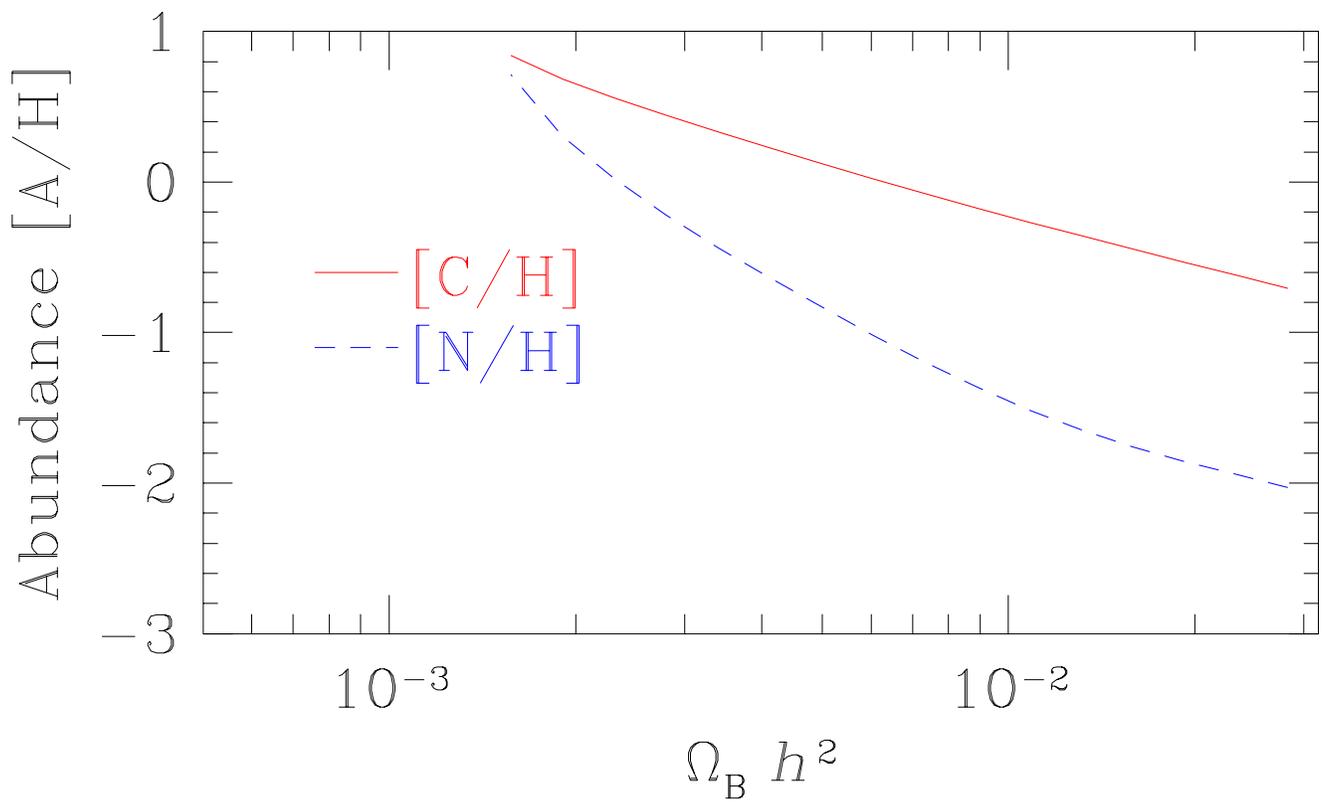
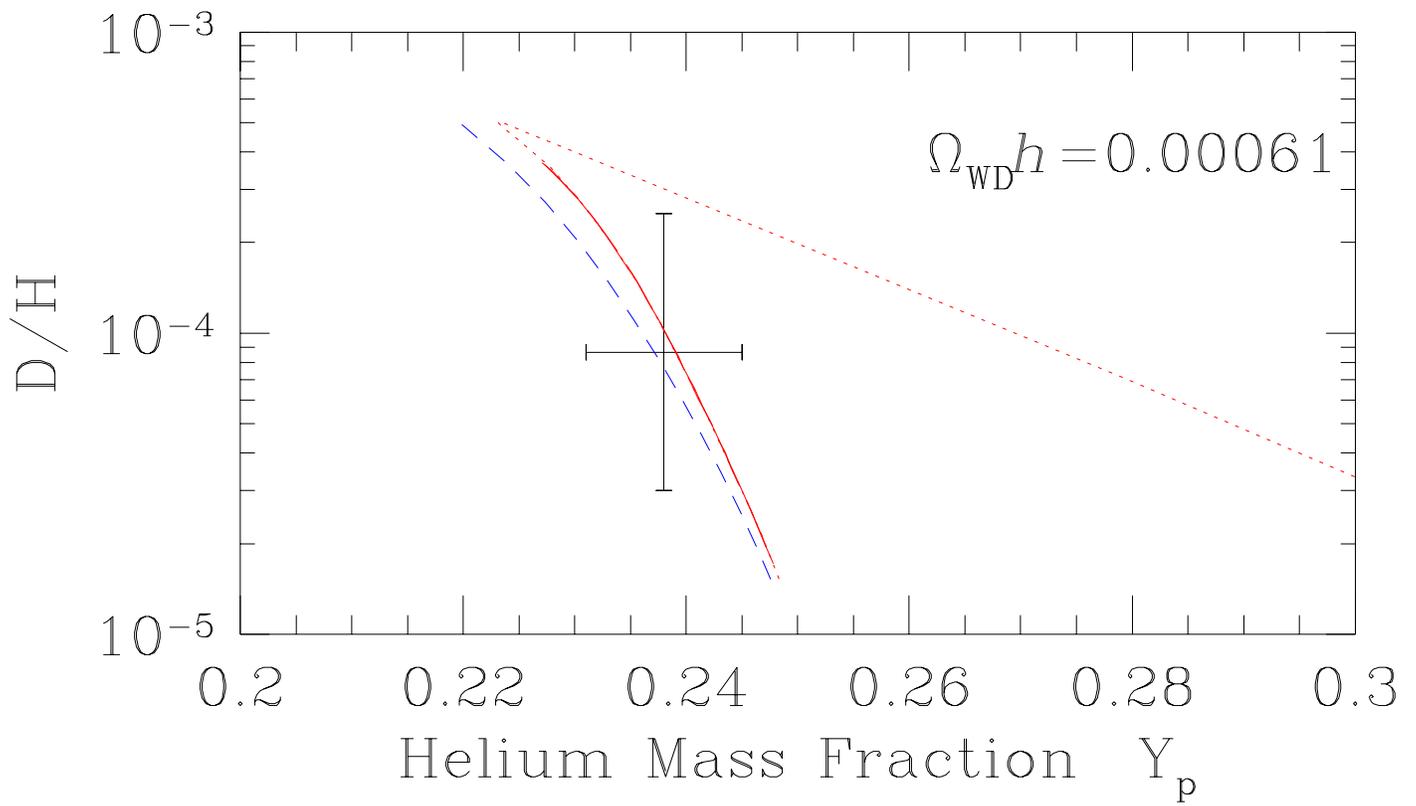
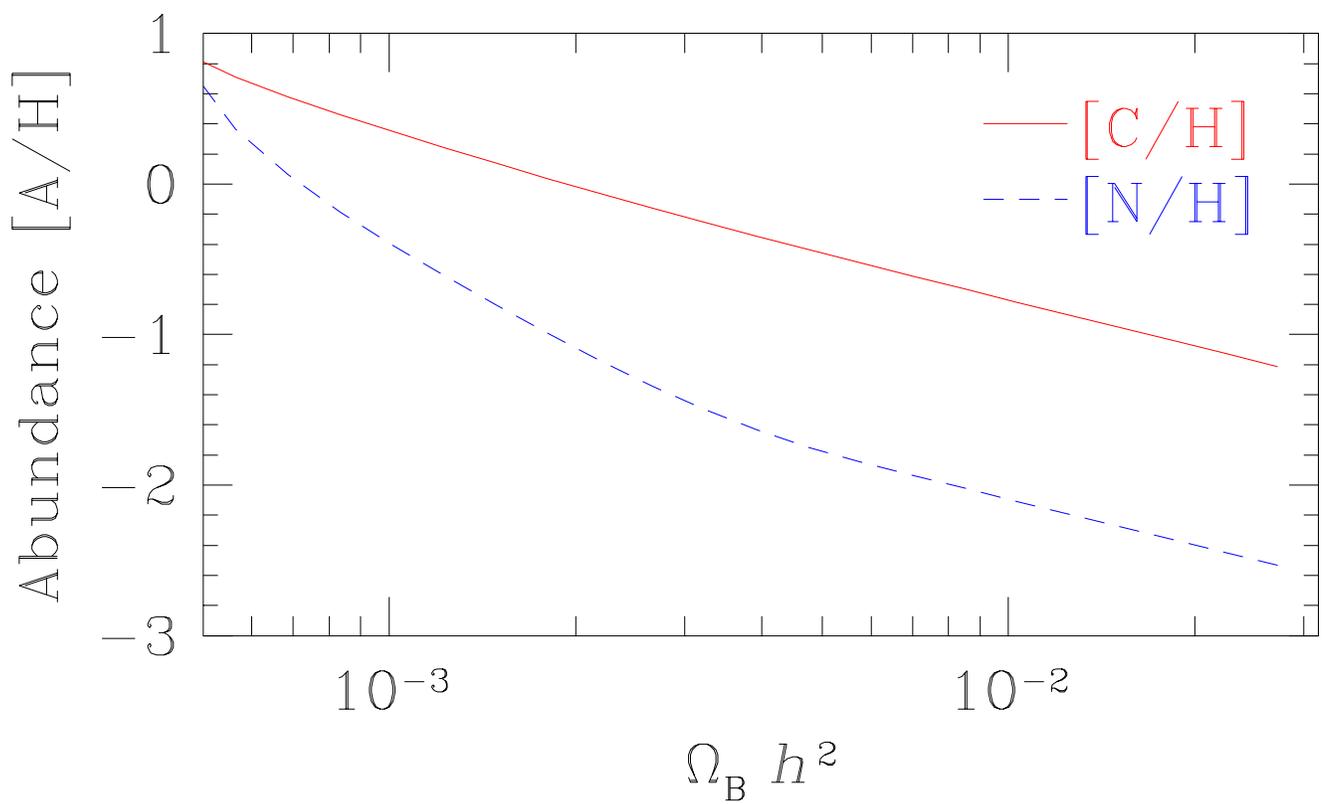
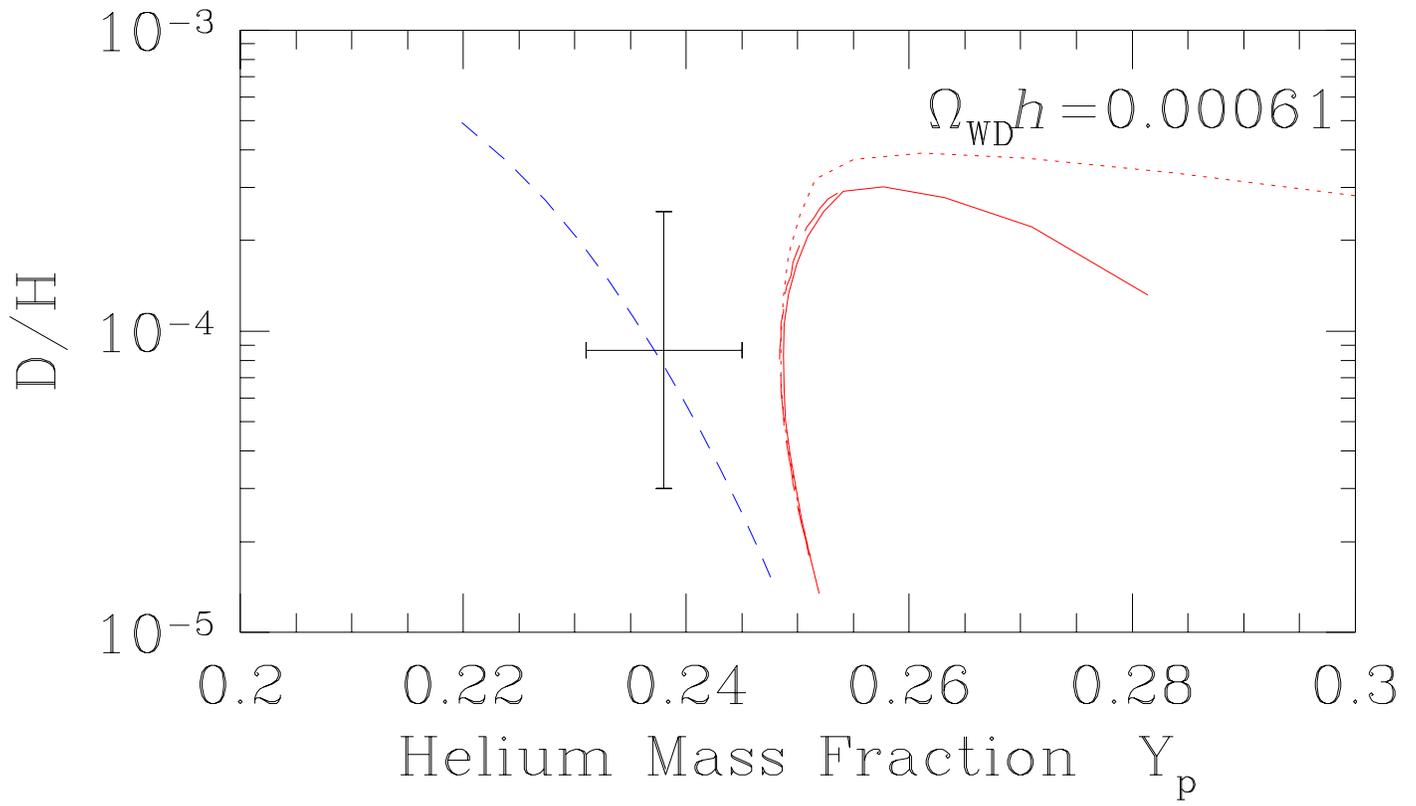
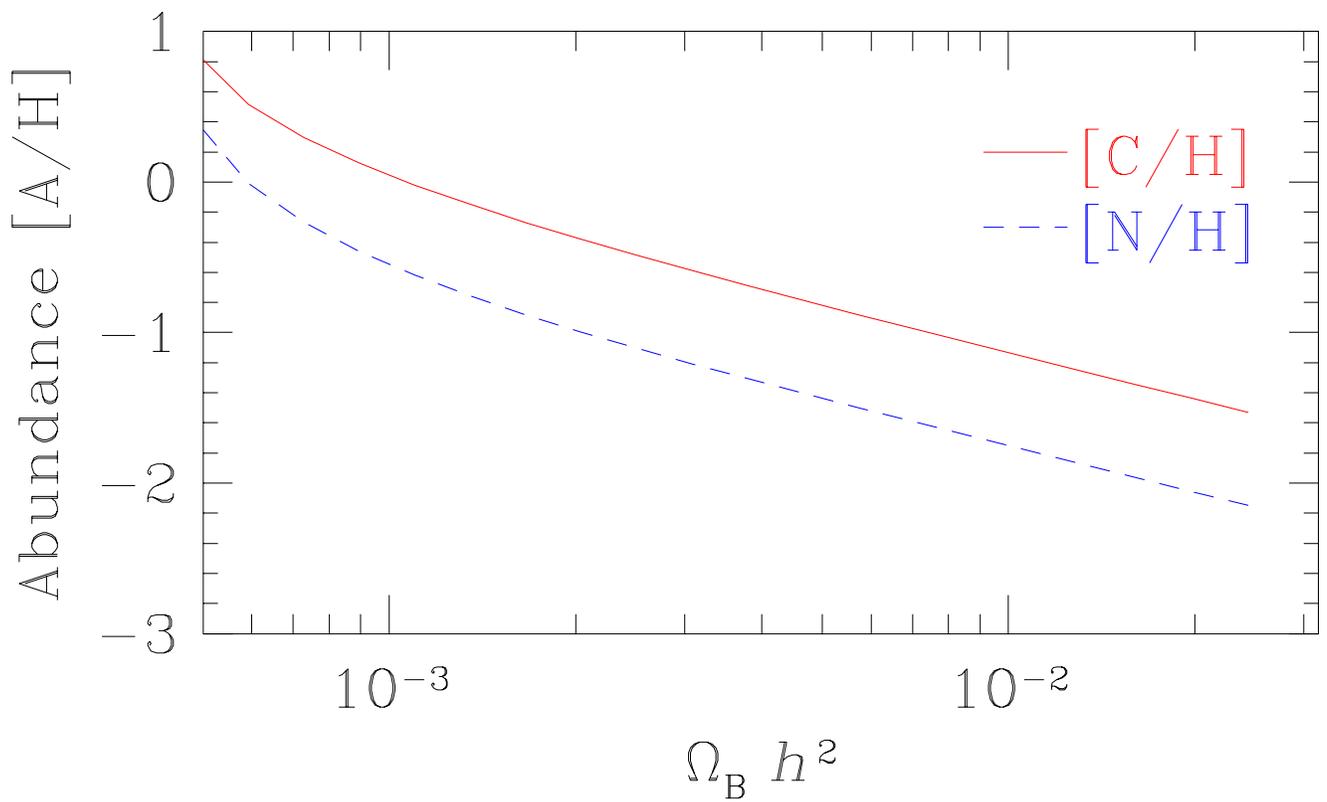
\end{document}